\documentclass[journal]{IEEEtran}

\usepackage{subfig}
\usepackage{float}
\usepackage{algorithm}
\usepackage{algpseudocode}
\usepackage{graphicx}
\usepackage{amsmath,amssymb,amsfonts}
\usepackage{url}
\usepackage{xcolor}
\usepackage{hyperref}
\usepackage[normalem]{ulem}
\usepackage{flushend}
\usepackage{stfloats}
\usepackage[final,expansion=true,protrusion=true,tracking=false]{microtype}

\newif\ifdiffmarkup
\diffmarkupfalse

\newcommand{\add}[1]{\ifdiffmarkup\textcolor{blue}{#1}\else#1\fi}
\newcommand{\del}[1]{\ifdiffmarkup\textcolor{red}{\sout{#1}}\fi}
\newcommand{\addblock}{\ifdiffmarkup\color{blue}\fi}


\begin{document}

\title{Comb Test: Histogram Uniformity Testing Based on Discrete Total Variation}

\author{Nikola Bani{\'{c}} and Neven Elezovi{\'{c}}
}

\markboth{IEEE Signal Processing Letters}
{Bani{\'{c}} and Elezovi{\'{c}}: Comb Test: Histogram Uniformity Testing Based on Discrete Total Variation}
\maketitle

\renewcommand\footnoterule{{\hrule height 0pt}}
\let\thefootnote\relax\footnotetext{
Manuscript received. Both authors contributed equally to this work.

Nikola Bani{\'{c}} is with Gideon Brothers, 10000 Zagreb, Croatia (email: nbanic@gmail.com). Neven Elezovi{\'{c}} is with Department of Applied Mathematics, Faculty of Electrical Engineering and Computing, University of Zagreb, 10000 Zagreb, Croatia (email: neven@element.hr). Code and data: \url{https://github.com/DiscreteTotalVariation/CombTest}.

Digital Object Identifier
}

\begin{abstract}
Histogram uniformity testing is a common statistical task usually performed using Pearson's chi-square test. \del{This paper proposes a new test based on the discrete total variation that is easy to compute and, for comb-like (alternating) deviations, achieves up to $67\%$ higher statistical power than Pearson's chi-square test, making it a complement to standard tests. The exact null distribution is computed via dynamic programming, and a gamma approximation with Monte Carlo estimation extends the test to arbitrarily large sample sizes. Experiments on simulated ADC alternating differential nonlinearity and on rounding bias detection in scientific data confirm the claims.}\add{This paper proposes a complementary test based on the discrete total variation of the bin counts, which is trivial to compute and, unlike chi-square, exploits bin ordering. Its exact null distribution follows from dynamic programming, and a gamma approximation, verified up to $N=5\cdot10^4$, extends it to large samples. Matching alternatives in $\ell_1$ distance locates the test: it gains over chi-square only in the period-2 comb direction and is least sensitive to smooth, single-jump, and depleted-block departures, for which chi-square is preferable. Experiments on converter nonlinearity, rounding bias, and real images confirm the analysis.} \del{The Python source code and precomputed data are}\add{Code and data are} available at~\url{https://github.com/DiscreteTotalVariation/CombTest}.
\end{abstract}


\IEEEpeerreviewmaketitle

\newcommand{\E}[1]{\mathbb{E}\left[ #1 \right]}
\newcommand{\A}[1]{\lvert #1 \rvert}
\newcommand{\Var}[1]{\mathrm{Var}\left[ #1 \right]}
\newcommand{\V}[1]{\left\lVert #1 \right\rVert_{V}}
\newcommand{\Norm}[2]{\left\lVert #1 \right\rVert_{#2}}
\newcommand{\N}[2]{\mathcal{N}\left(#1,#2\right)}

\section{Introduction}
\label{sec:introduction}

\IEEEPARstart{L}{et} $\mathbf{x}_n$ be a histogram with $n$ bins, where $x_i$ is the count of values from a sample of $N$ in the $i$-th bin so that
\begin{equation}
\label{eq:sum}
\sum_{i=1}^{n}x_i=N.
\end{equation}
Pearson's chi-square test~\cite{pearson1900chi} and the $G$-test~\cite{wilks1938large}, commonly used for uniformity testing, treat bins independently and thus cannot leverage the structure of alternating deviations. The discrete total variation~(DTV)~\cite{banic2020tvor} captures these:
\begin{equation}
    \label{eq:dtv}
    \V{\mathbf{x}_n}=\sum_{i=2}^{n}\lvert x_{i}-x_{i-1}\rvert.
\end{equation}
DTV is the $\ell_1$ norm of the \del{discrete first-order differences of the histogram; it is zero when all bins are equal and amplifies}\add{histogram's first-order differences: zero when all bins are equal, and large for} oscillatory deviations that global divergence measures average out. Unlike \del{Pearson's }chi-square\del{ test}, \del{DTV depends on bin ordering and is therefore suitable when bins have a natural order, such as ADC output codes, digit values, or pixel intensities encountered in imaging and signal processing.}\add{it depends on bin ordering, so it suits naturally ordered bins such as ADC codes, digit values, or pixel intensities. Testing uniformity on such bins is routine: IEEE~Std~1241~\cite{ieee1241} prescribes it for ADC code-occupancy validation, terminal-digit screening exposes rounding and fabrication artifacts in scientific and financial data~\cite{nigrini2012benford}, randomness suites rely on it~\cite{rukhin2010statistical}, and pixel histograms in image forensics and steganalysis~\cite{fridrich2009steganography}.}

Anderson--Darling~\cite{anderson1954test}, Kolmogorov--Smirnov~\cite{massey1951ks}, and Cram\'{e}r--von Mises~\cite{cramer1928composition} target continuous distributions; Neyman's smooth tests~\cite{neyman1937smooth}, Wald--Wolfowitz~\cite{wald1940test}, and von Neumann~\cite{von1941distribution} target smooth, binary-sequence, and serial-correlation alternatives\del{ (the latter two operate on raw sequences rather than histograms, so they do not directly apply when only bin counts are available)}\add{, the last two on raw sequences, though their histogram-count analogues serve as baselines in Section~\ref{subsec:where}}. The Cressie--Read family~\cite{cressie1984multinomial} generalizes \del{chi-square and $G$-test but remains order-agnostic. None targets comb-like deviations where $p_i = 1/n + (-1)^i\delta$ for $\delta > 0$, which arise in practice from ADC capacitor mismatch and numerical rounding in scientific data.}\add{both chi-square and the $G$-test but is order-agnostic. Modern discrete distribution testing concerns sample complexity and minimax risk over large alphabets~\cite{canonne2022survey}: coincidence-based tests are optimal in the sparse regime~\cite{paninski2008coincidence}, minimax risk is characterized for smooth and inhomogeneous alternatives~\cite{ingster2003nonparametric,lepski1999minimax} and, over large alphabets, for missing-ball ones~\cite{kipnis2026minimax}, and histogram-structured distributions admit nearly tight bounds~\cite{canonne2022histogram}. Those classes---smooth departures and depleted blocks---are, as Table~\ref{tab:l1} shows, exactly where DTV is weakest, so this work complements rather than competes: it addresses the opposite, highest-frequency end, comb-like deviations $p_i = 1/n + (-1)^i\delta$ arising from ADC capacitor mismatch and numerical rounding, and it supplies an \emph{exact finite-sample} null, not an asymptotic rate, as per-device ADC characterization at small~$N$ requires.}

\del{This paper characterizes the DTV distribution under the uniform null hypothesis, yielding the \textbf{comb~test}~(CT), named after the targeted alternative shape (unrelated to comb filters). Section~\ref{sec:exact} presents exact computation via dynamic programming, Section~\ref{sec:approximate} a gamma approximation~\cite{johnson1995continuous} for large~$N$, Section~\ref{sec:results} confirms higher power than chi-square and $G$-test for comb distributions, and Section~\ref{sec:conclusions} concludes the paper and discusses directions for future research.}\add{TVOR~\cite{banic2020tvor} used DTV to \emph{rank} histograms by irregularity against a smooth model fitted across a collection: an outlier score with no null distribution or $p$-value. Here we derive its distribution under the uniform multinomial null for a \emph{single} histogram, exactly and for finite~$N$, turning DTV into a test with a valid false positive rate (FPR), and map where it is and is not powerful. The result is the \textbf{comb~test}~(CT), named after its targeted shape (unrelated to comb filters).}

\section{Calculating the exact DTV distribution}
\label{sec:exact}

Under the uniform null hypothesis $H_0$, each of $N$ values independently falls into one of $n$ equiprobable bins (multinomial model), yielding $n^N$ equally likely ordered outcomes. The count for each DTV value \del{can be computed via}\add{follows from} dynamic programming~\cite{bellman2015applied}\add{ that builds the histogram one bin at a time, enumerating all values for each new bin and extending the DTV counts}. Maximum DTV is $2N$ for $n>2$ (\del{attained when }all $N$ values \del{concentrate }in non-adjacent bins), $N$ for $n=2$, and $0$ for $n=1$.\del{ The algorithm builds the histogram one bin at a time, enumerating all possible values for each new bin and extending the DTV counts at each successive step.}

\subsection{Calculation specifics}
\label{subsec:specifics}

Let $C(i, M, m, d)$ be the number of ways to place $M$ of the $N$ \add{distinguishable }values into the first $i$ bins with $m$ values in the $i$-th bin and DTV $d$\add{ over those bins alone}, where $1\leq i\leq n$, $0\leq M\leq N$, $0\leq m\leq M$, $0\leq d\leq 2N$. $m$ is tracked since appending a bin with value $m'$ increases the DTV by $\A{m'-m}$. Counts for $i=1$ yield those for $i=2$ by enumerating all \del{possible }next-bin values, and so on\del{. The}\add{; the} \del{result }\add{number $D_{N,n}(d)$ of outcomes with DTV $d$ }sums over all last-bin values:
\begin{equation}
    \label{eq:dnnd}
    D_{N,n}(d)=\sum_{m=0}^{N}C(n, N, m, d).
\end{equation}

\subsubsection{Single bin}
\label{subsubsec:single}

For $i=1$, the only valid state has $m=M$ (all $M$ values must be in the sole bin) with $d=0$, giving $C(1, M, M, 0) = \binom{N}{M}$ and $C(1, M, m, d) = 0$ otherwise.

\subsubsection{Added bins}
\label{subsubsec:added}

For $i > 1$, $C(i, M, m, d)$ follows from \del{the $i-1$ case}\add{case $i-1$}. \del{Since}\add{With} $m$ values \del{go into}\add{in} bin $i$ and $M-m$ \del{into}\add{in} the first $i-1$ bins, the previous bin \del{had value}\add{value $k$ ranges over} $0 \leq k \leq M - m$. Appending \del{value }$m$ after $k$ \del{increases}\add{raises} the DTV by $\A{m-k}$, so only counts with prior DTV $d - \A{m-k}$ contribute\del{.}\add{; $C$ is zero whenever an argument leaves its stated range, e.g.\ for $d-\A{m-k}<0$, so Section~\ref{subsubsec:single} is the only initialization needed.} The factor $\binom{N-M+m}{m}$ counts \del{ways to choose}\add{choices of} $m$ \del{of}\add{from} the $N-(M-m)$ values \del{not in bins $1$ through $i-1$}\add{outside bins $1,\ldots,i-1$}. Summing over $k$:
\begin{equation}
    \label{eq:dp}
    \begin{gathered}
    C(i, M, m, d) \\
    =\binom{N-M+m}{m}\cdot\sum_{k=0}^{M-m}C(i-1, M-m, k, d-\A{m-k}).
    \end{gathered}
\end{equation}
\del{The complexity is $O(nN^4)$ time and $O(N^3)$ memory (only bin $i{-}1$ is needed for bin $i$); the implementation uses arbitrary-precision integers to ensure exact computation. Exact computation is feasible for $N$ up to several hundred; beyond that, the gamma approximation (Section~\ref{sec:approximate}) is used. Once precomputed, evaluating CT costs $O(n)$ per test---identical to chi-square and the $G$-test. MC precomputation ($K$ multinomial samples plus a discrete Cram\'{e}r--von Mises~\cite{cramer1928composition} fit) typically runs in under one second for $K{=}50{,}000$, $n{=}10$; a~C implementation (GMP) is provided in the repository. Section~\ref{subsec:algorithm} uses a different iteration strategy for efficiency.}\add{Counting states and transitions gives the complexity: for each of the $n$ bins, $M$, $m$, and $d$ each take $O(N)$ values, hence $O(N^3)$ states, each a sum of $O(N)$ terms, so $O(nN^4)$ time; as bin $i$ needs only layer $i{-}1$, two layers suffice, i.e.\ $O(N^3)$ memory. The counts sum to $n^N$, outgrowing machine integers, so arbitrary-precision arithmetic keeps them exact, feasible for $N$ to a few hundred; beyond that Section~\ref{sec:approximate}'s gamma approximation is used. Once precomputed for a given $N$ and $n$, CT costs $O(n)$ per test, like chi-square and $G$.} The probability of \del{DTV value $d$}\add{DTV $d$} under $H_0$\del{ is}\add{:}
\begin{equation}
    \label{eq:p}
    p_{N, n}^{(d)}=\frac{D_{N,n}(d)}{n^N}.
\end{equation}

\subsection{Implementation}
\label{subsec:algorithm}

Algorithm~\ref{alg:p} implements \del{the procedure from }Section~\ref{subsec:specifics}\add{ with a faster iteration order}. The array $\text{cc}(s, l, d)$ counts \del{the ways to place}\add{placements of} $s$ values into the current bins with last bin value $l$ and DTV~$d$. \del{Instead of}\add{Rather than} summing as in Eq.~\ref{eq:dp}, it skips zero entries (line~\ref{alg:skip}) and \del{distributes contributions to all valid new-bin values}\add{pushes contributions to valid new bins} (line~\ref{alg:newbin}).

\begin{algorithm}
\fontsize{6.2}{7.0}\selectfont
\caption{Calculating the exact DTV distribution}
\label{alg:p}
\hspace*{\algorithmicindent}\textbf{Input:} $N$, $n$ \\
\hspace*{\algorithmicindent}\textbf{Output:} $p_{N, n}^{(0)}$, $p_{N, n}^{(1)}$, $\ldots$, $p_{N, n}^{(2N)}$
\begin{algorithmic}[1]
\State cc $\gets$ Zeros($N+1$, $N+1$, $2N+2$) \Comment{current counts}
\For{$i\in\{0, 1, 2, \ldots, N\}$} \Comment{initialize the first bin's counts}
    \State cc$(i, i, 0)\gets \binom{N}{i}$
\EndFor
\For{$b\in\{2, \ldots, n\}$}
    \State pc $\gets$ cc \Comment{earlier current counts become previous}
    \State cc $\gets$ Zeros($N+1$, $N+1$, $2N+2$) \Comment{clear the values}
    \For{ps $\in\{0, \ldots, N\}$} \Comment{previous sum}
        \State r $\gets$ $N$ $-$ ps \Comment{remaining values}
        \For{pl $\in\{0, \ldots, \text{ps}\}$} \Comment{previous last bin value}
            \For{p\_dtv $\in\{0, \ldots, 2\cdot\text{ps}\}$} \Comment{previous DTV}
                \State prev $\gets$ pc(ps, pl, p\_dtv)
                \If{prev$>0$}\label{alg:skip}\Comment{important for speed}
		              \For{nl $\in\{0, \ldots, \text{r}\}$}\label{alg:newbin} \Comment{new bin value}
                        \State ns $\gets$ ps $+$ nl \Comment{new sum}
                        \State n\_dtv $\gets$ p\_dtv $+ |$nl $-$ pl$|$
                        \State cc(ns, nl, n\_dtv) $\gets$ cc(ns, nl, n\_dtv) $+$ prev $\cdot$ $\binom{\text{r}}{\text{nl}}$
                    \EndFor
                \EndIf
            \EndFor
        \EndFor
    \EndFor
\EndFor
\For{$d\in\{0, 1, 2, \ldots, 2N\}$}
    \State $D_{N,n}(d)\gets\sum_{m=0}^{N}\text{cc}(N, m, d)$
    \State $p_{N, n}^{(d)}\gets\frac{D_{N,n}(d)}{n^N}$
\EndFor
\end{algorithmic}
\end{algorithm}

\subsection{Statistical test}
\label{subsec:test}

\del{For observed DTV value $d$, the}\add{The} $p$-value \add{for DTV $d$ }is
\begin{equation}
\label{eq:p_value}
p=1-\sum_{i=0}^{d-1}p_{N, n}^{(i)},
\end{equation}
\del{where the empty sum ($d=0$) yields}\add{empty for $d=0$, so} $p=1$\del{ by convention}.
\del{As Section~\ref{sec:results} shows, CT outperforms Pearson's chi-square for comb-shaped histograms. CT is one-sided: it rejects only when DTV is unusually large. A two-sided variant (also detecting unusually small DTV, e.g., overly regular allocations) is left for future work.}\add{CT rejects only large DTV; a two-sided variant for overly regular histograms is future work.}

\begin{figure}[!t]
    \centering
	\subfloat[]{%
	\includegraphics[width=\linewidth]{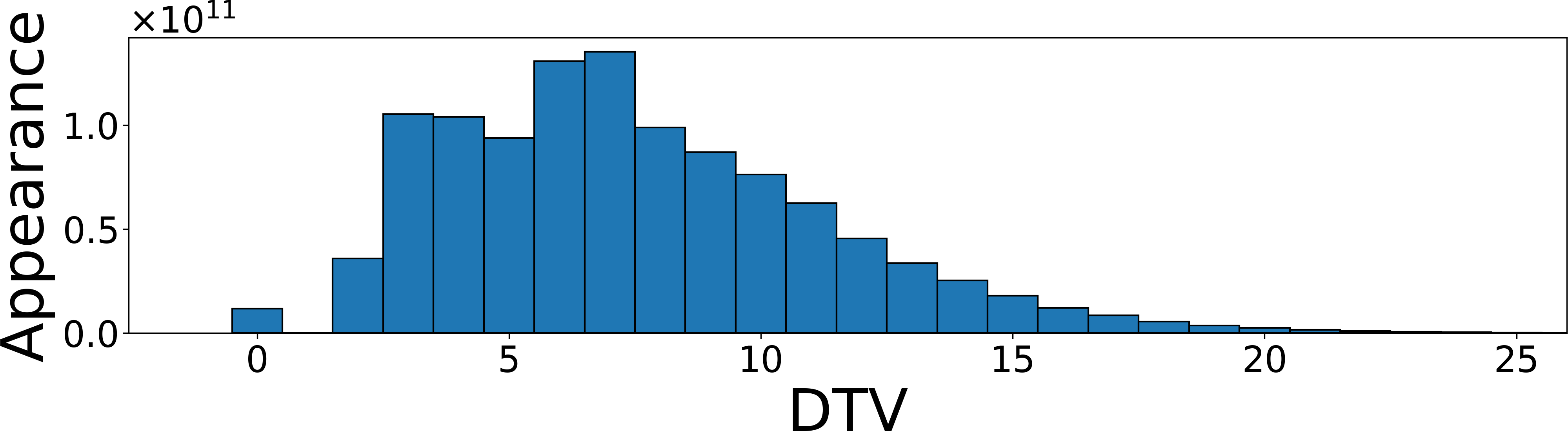}%
	\label{fig:dtv_N_20_n_4}%
	}
	\\
	\subfloat[]{%
	\includegraphics[width=\linewidth]{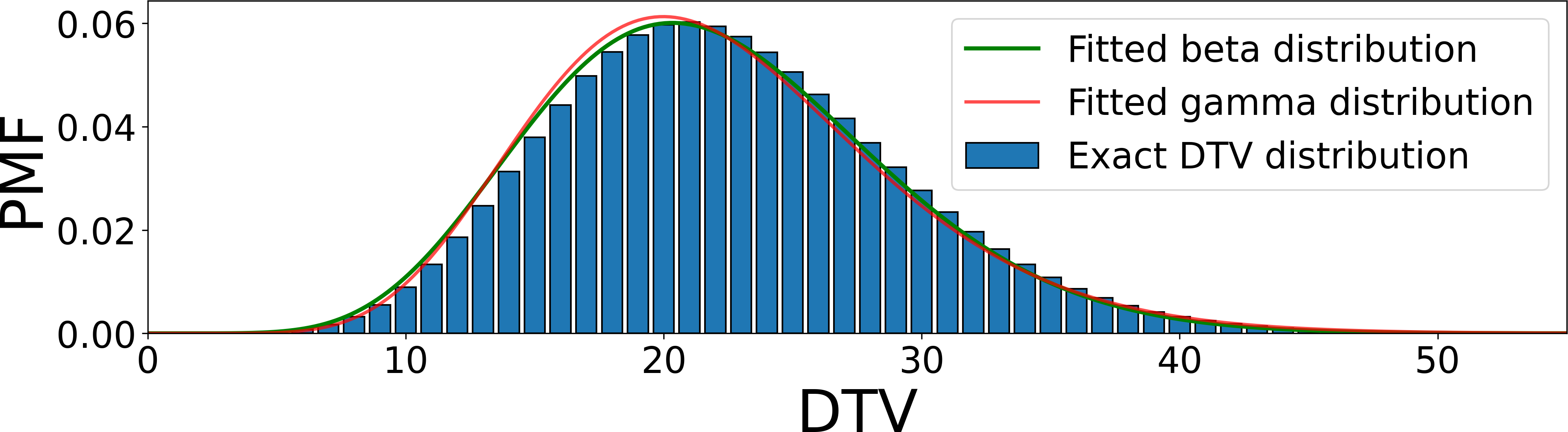}%
	\label{fig:dtv_N_50_n_10_and_beta}%
	}
    \caption{DTV distribution under \del{the uniform null hypothesis for}\add{$H_0$:} a)~$N=20$, $n=4$\del{ and}\add{;} b)~$N=50$, $n=10$\del{. In~(b), fitted beta and gamma distributions are overlaid.}\add{, with beta and gamma fits overlaid.}}
	\label{fig:dtv}
\end{figure}

\begin{figure}[!t]
    \centering
    \includegraphics[width=\linewidth]{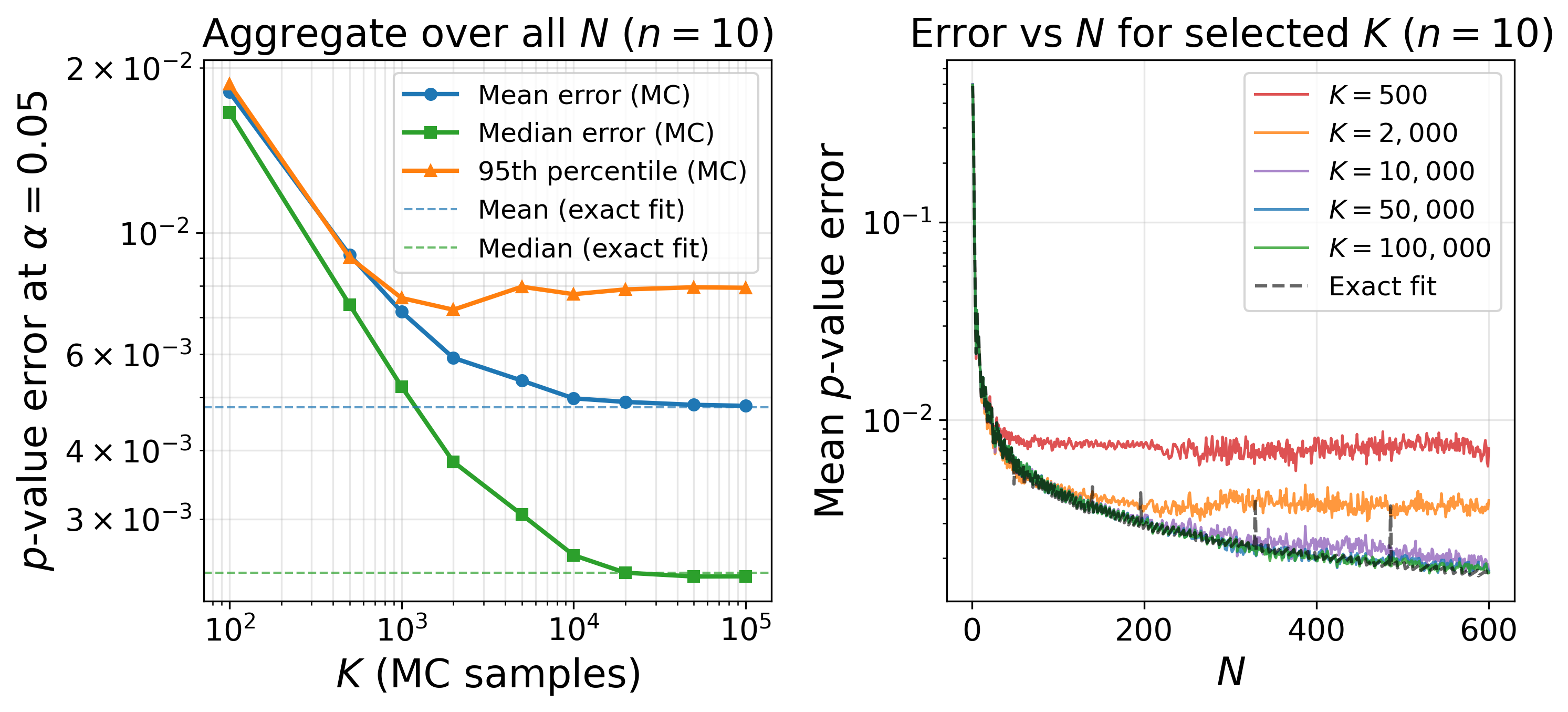}
    \caption{MC convergence\del{ for}\add{,} $n=10$, \del{$N=1,\ldots,600$ ($100$ repeats per $K$)}\add{$N\leq600$, $100$ repeats/$K$}. a)~\del{$p$-value }error vs.\ $K$\del{; dashed: exact-fit baseline. b)~Error vs.\ $N$; larger $K$ collapse onto the exact-fit curve}\add{; b)~vs.\ $N$, larger $K$ nearing the exact fit}.}
    \label{fig:mc_convergence}
\end{figure}

\begin{figure*}[!t]
    \centering
	\subfloat[]{
	\includegraphics[width=0.32\linewidth]{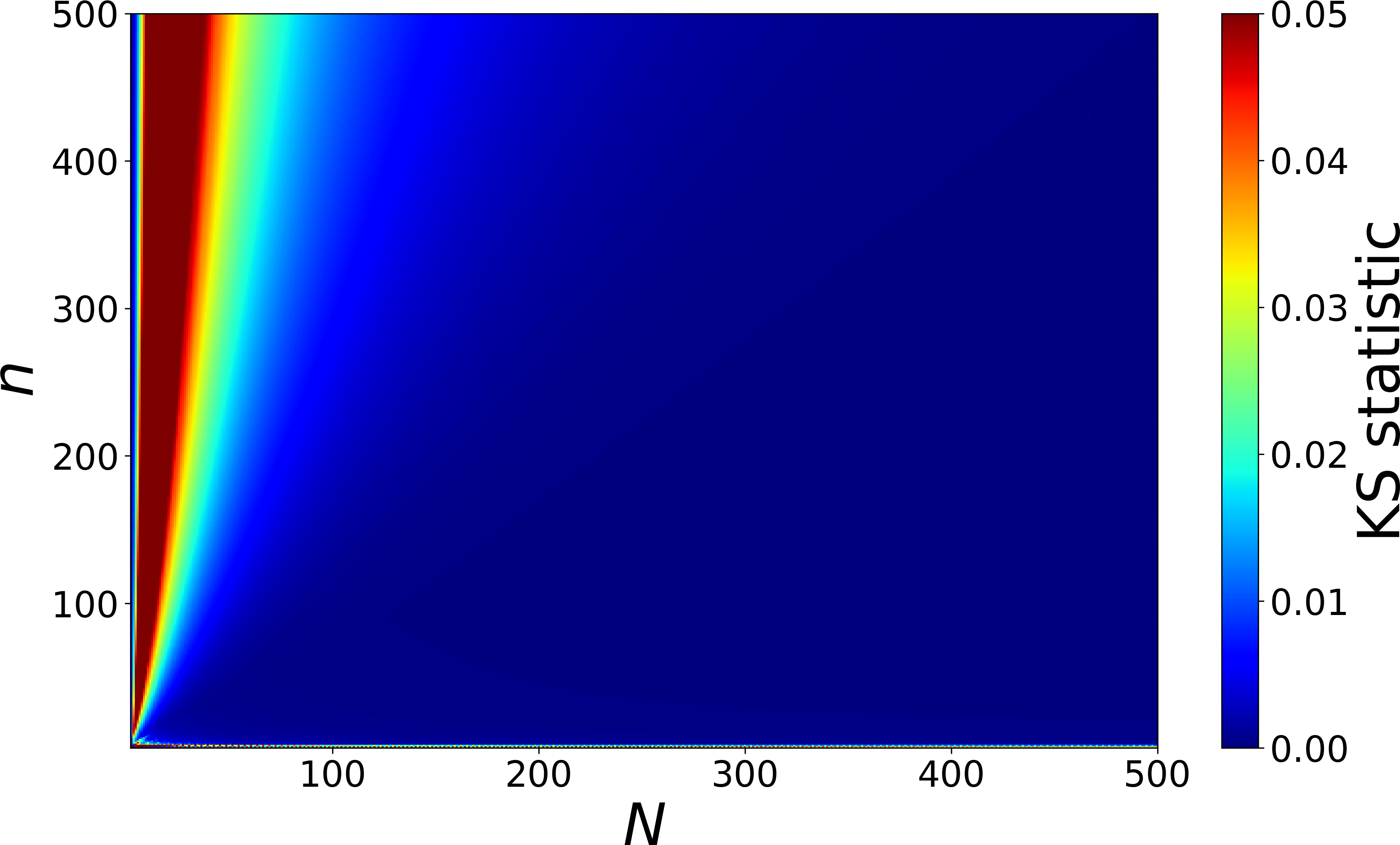}
	\label{fig:heatmap_full}
	}
	\subfloat[]{
	\includegraphics[width=0.32\linewidth]{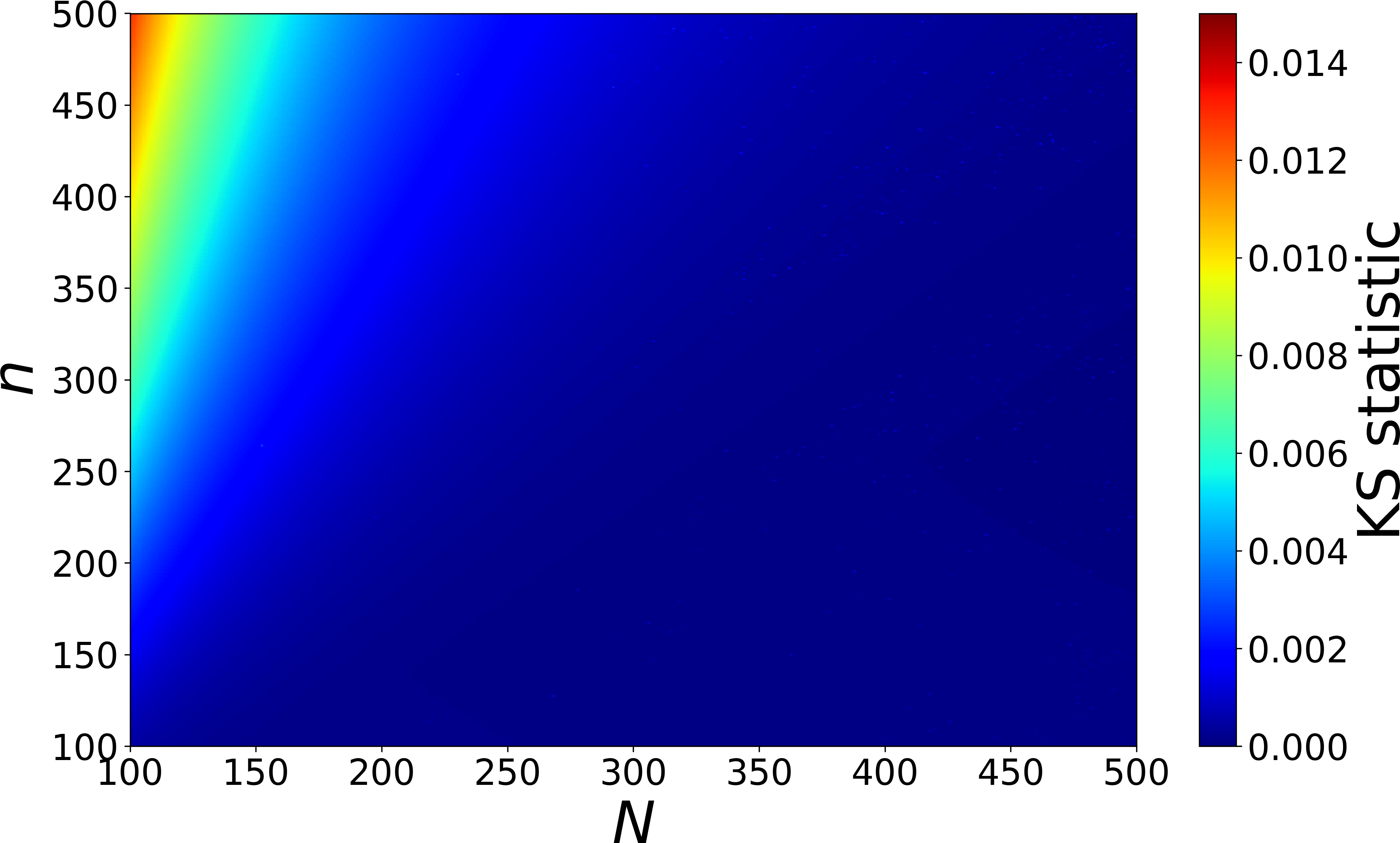}
	\label{fig:heatmap_zoomed}
	}
	\subfloat[]{
	\includegraphics[width=0.32\linewidth]{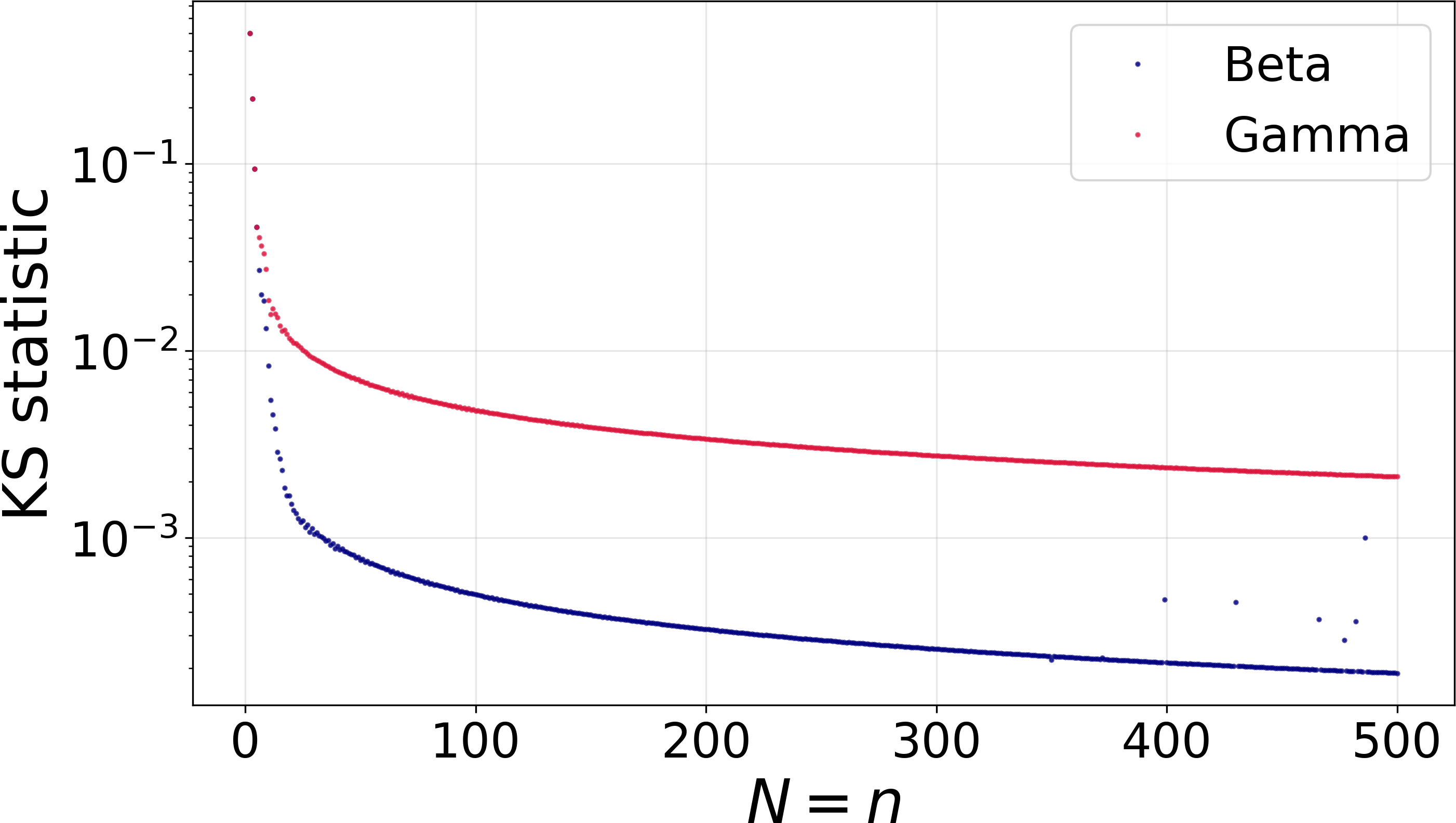}
	\label{fig:ks_diagonal}
	}
    \caption{KS statistic\del{ between exact and approximated DTV distributions. a)~Beta fit,}\add{, exact vs.\ approximated DTV CDFs. a)~Beta,} $N, n \in \{2, \ldots, 500\}$; \del{the $n > N$ region shows large error due to degenerate distributions. b)~Beta fit,}\add{large only where $n > N$. b)~Beta,} $N, n \in \{100, \ldots, 500\}$, \del{revealing fine structure}\add{fine structure}. c)~\del{Along the diagonal }$N{=}n$: \del{both }beta (blue) and gamma (red) \del{improve with increasing $N$, but the beta fit achieves consistently lower KS error.}\add{both improve with $N$, beta the more.}}
	\label{fig:goodness_of_fit_beta}
\end{figure*}

\begin{figure*}[!t]
    \centering
	\subfloat[]{
	\includegraphics[width=0.48\linewidth]{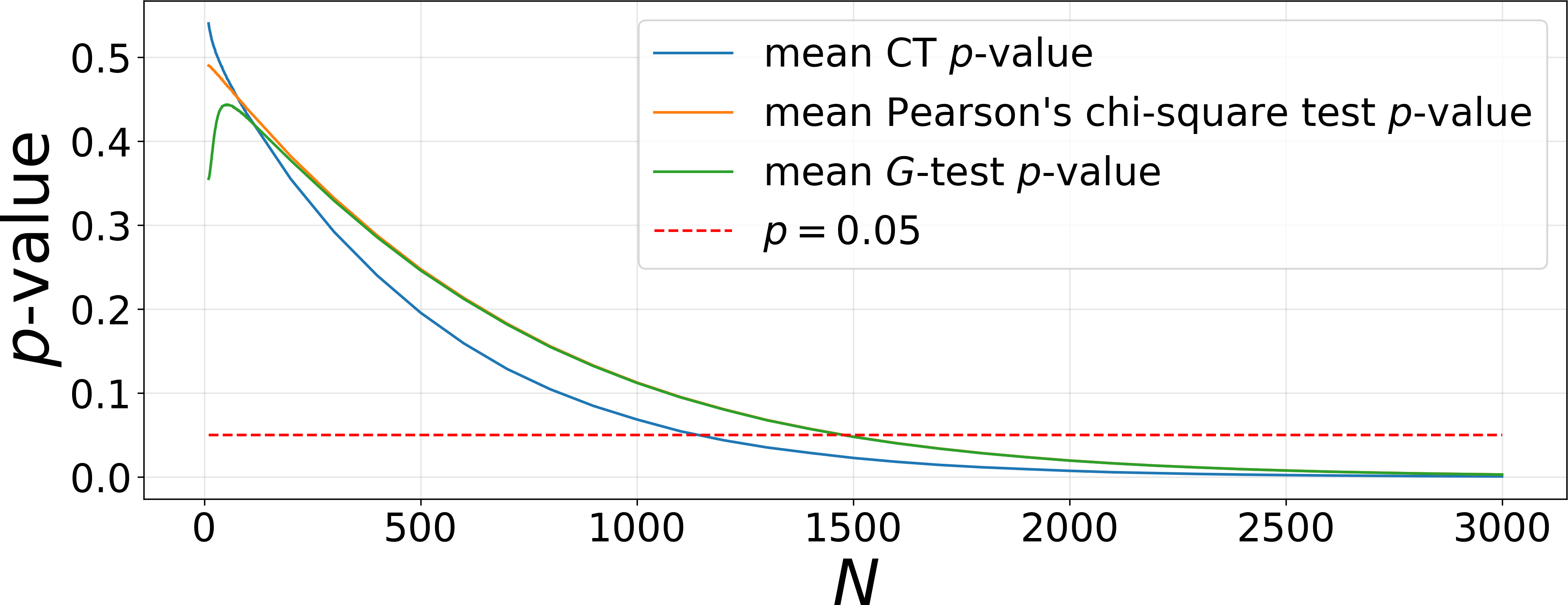}
	\label{fig:mean_p_values_d_0_step_100}
	}
	\subfloat[]{
	\includegraphics[width=0.48\linewidth]{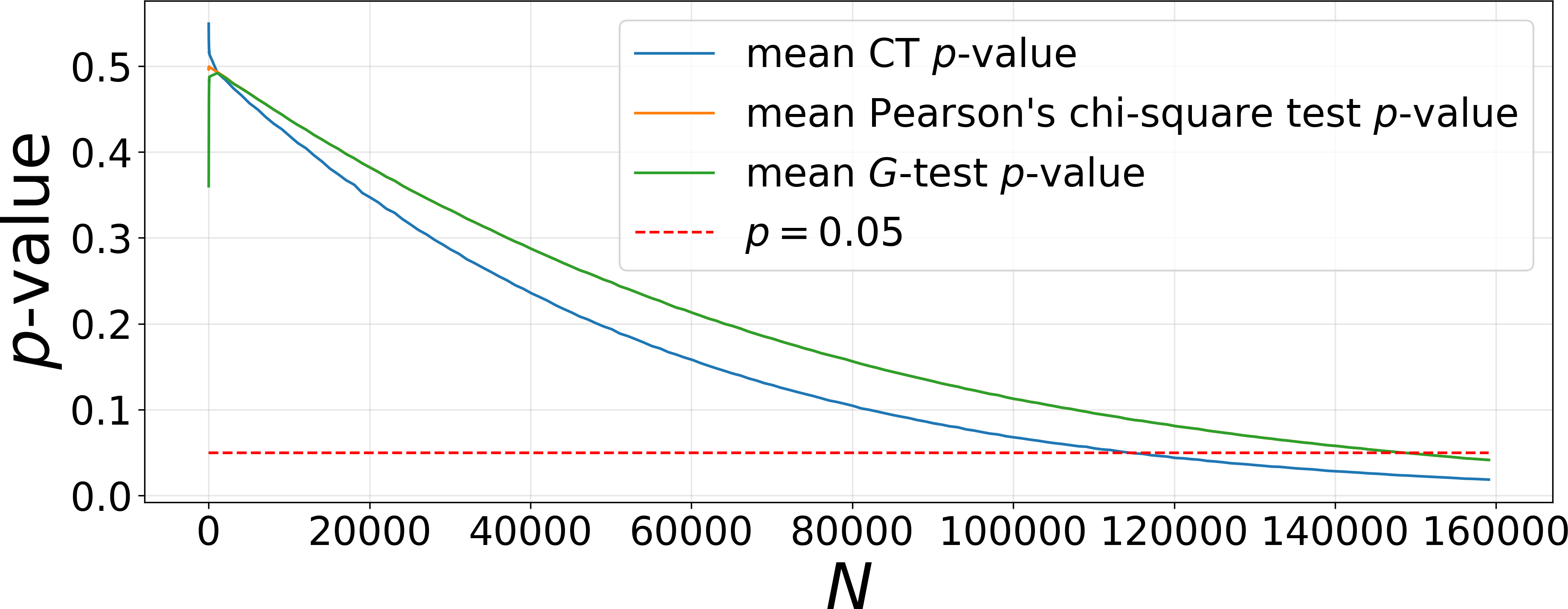}
	\label{fig:mean_p_values_d_1_step_1000}
	}
    \caption{Mean $p$-value\del{ for}\add{,} last-digit uniformity\del{ (}\add{, }$10^6$ histograms\del{ per }\add{/}$N$\del{)}, $r=1$~(a)\del{ and}\add{,} $r=2$~(b) decimals\del{. Note different $x$-axis scales ($r=2$ effect is ${\sim}100\times$ weaker). Lower is better.}\add{; note the differing $x$-axis scales.}}
	\label{fig:mean_p_values}
\end{figure*}

\begin{figure}[!b]
    \centering
    \includegraphics[width=\linewidth]{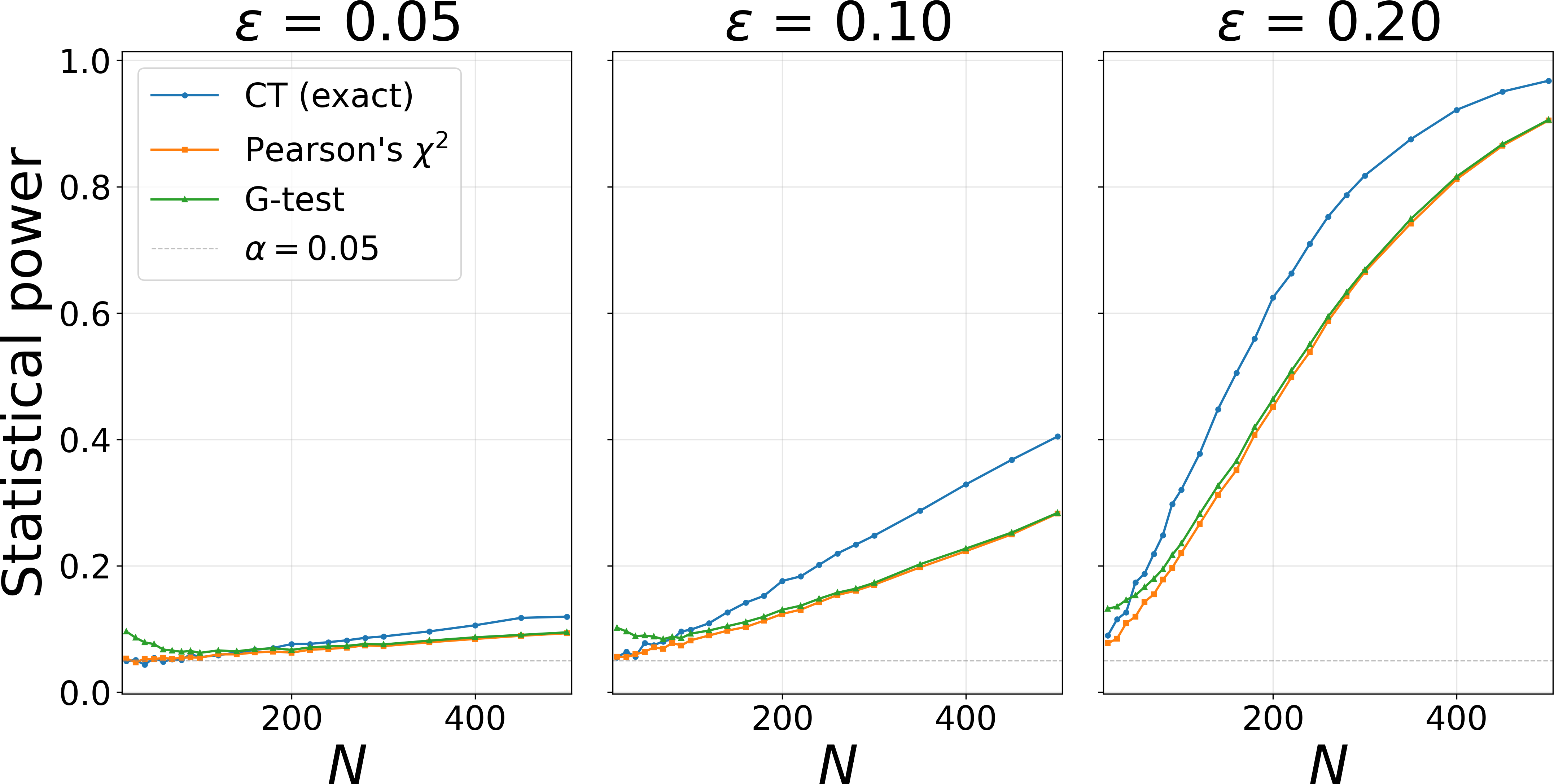}
    \caption{Power for detecting alternating DNL in a simulated ADC ($n=10$, $\varepsilon \in \{0.05, 0.10, 0.20\}$). \del{CT outperforms chi-square and $G$-test. }$50{,}000$ trials, $\alpha=0.05$.}
    \label{fig:adc_power}
\end{figure}

\begin{table*}[!t]
\addblock
\centering
\caption{\footnotesize \add{Power at $\alpha=0.05$; all alternatives at $\ell_1$ distance $0.2$ from uniformity, so only the \emph{shape} differs. AC: alternating contrast (oracle for the period-2 comb); vN: von Neumann ratio; RUN: runs test; CO: coincidence count~\cite{paninski2008coincidence}; exact nulls except vN (MC); valid but conservative. Row~1: attained FPR. $-$: $G$ invalid at $n=256$ (FPR $0.457$), CO vacuous at $n=10$ (all bins occupied). $50{,}000$/$20{,}000$ trials for $n=10$/$256$, SE ${\leq}\,0.004$; bold: best of CT, $\chi^2$, $G$, CO.}}
\label{tab:l1}
\fontsize{6.4}{7.2}\selectfont
\setlength{\tabcolsep}{3.2pt}
\begin{tabular}{l|ccccccc|ccccccc}
\hline
& \multicolumn{7}{c|}{$n=10$, $N=200$} & \multicolumn{7}{c}{$n=256$, $N=500$} \\
Departure from uniformity & CT & $\chi^2$ & $G$ & AC & vN & RUN & CO & CT & $\chi^2$ & $G$ & AC & vN & RUN & CO \\
\hline
none (attained FPR) & 0.048 & 0.049 & 0.052 & 0.040 & 0.049 & 0.035 & $-$ & 0.045 & 0.049 & 0.457 & 0.043 & 0.053 & 0.045 & 0.034 \\
comb, period $2$ & \textbf{0.623} & 0.449 & 0.461 & 0.786 & 0.370 & 0.184 & $-$ & \textbf{0.376} & 0.229 & $-$ & 0.994 & 0.321 & 0.136 & 0.123 \\
comb, period $4$ & 0.248 & 0.472 & \textbf{0.496} & 0.040 & 0.005 & 0.003 & $-$ & 0.140 & \textbf{0.221} & $-$ & 0.043 & 0.039 & 0.034 & 0.121 \\
comb, period $6$ & 0.184 & 0.471 & \textbf{0.495} & 0.190 & 0.008 & 0.010 & $-$ & 0.096 & \textbf{0.224} & $-$ & 0.301 & 0.018 & 0.044 & 0.119 \\
sinusoid, period $4$ & 0.579 & 0.728 & \textbf{0.752} & 0.068 & 0.002 & 0.001 & $-$ & 0.321 & \textbf{0.511} & $-$ & 0.044 & 0.026 & 0.028 & 0.299 \\
sinusoid, period $n$ & 0.059 & 0.534 & \textbf{0.551} & 0.040 & 0.001 & 0.022 & $-$ & 0.038 & \textbf{0.288} & $-$ & 0.047 & 0.002 & 0.129 & 0.157 \\
monotone ramp & 0.053 & 0.590 & \textbf{0.606} & 0.074 & 0.001 & 0.331 & $-$ & 0.035 & \textbf{0.310} & $-$ & 0.044 & 0.002 & 0.144 & 0.166 \\
single jump & 0.076 & 0.449 & \textbf{0.462} & 0.071 & 0.003 & 0.307 & $-$ & 0.039 & \textbf{0.227} & $-$ & 0.047 & 0.004 & 0.096 & 0.122 \\
depleted block & 0.130 & 0.731 & \textbf{0.795} & 0.040 & 0.001 & 0.093 & $-$ & 0.036 & \textbf{0.311} & $-$ & 0.042 & 0.001 & 0.197 & 0.214 \\
single spike & 0.674 & \textbf{0.876} & 0.840 & 0.315 & 0.026 & 0.010 & $-$ & 0.948 & \textbf{1.000} & $-$ & 0.587 & 0.000 & 0.059 & 0.441 \\
unstructured & 0.764 & 0.858 & \textbf{0.919} & 0.546 & 0.117 & 0.073 & $-$ & 0.240 & \textbf{0.396} & $-$ & 0.048 & 0.044 & 0.044 & 0.237 \\
\hline
\end{tabular}
\end{table*}

\section{Approximating the DTV distribution}
\label{sec:approximate}

\subsection{Distribution fitting}
\label{subsec:beta}

Exact computation is infeasible for large $N$, requiring a continuous approximation~\cite{rees2018essential}. \add{The recommended procedure is the \emph{gamma} approximation; beta is reported only as the best global fit. }Since \del{the }DTV \del{distribution }is bounded ($0 \leq \text{DTV} \leq 2N$ for $n > 2$) and right-skewed (Fig.~\ref{fig:dtv}), \del{the beta distribution is a natural fit, validated empirically in Section~\ref{subsec:beta_quality}.}\add{a bounded two-parameter family is natural.} Twenty-nine candidates\footnote{\del{Including beta, gamma, normal, Weibull, lognormal, and 24 others; see the repository for the full list.}\add{Beta, gamma, normal, Weibull, lognormal, and 24 more; see the repository.}}, \del{fitted by}\add{fitted against exact CDFs for $N, n \in \{2, \ldots, 500\}$ by} minimizing a discrete Cram\'{e}r--von Mises statistic \del{$W^2 = \sum_i (F_{\text{exact}}(x_i) - F_{\text{approx}}(x_i))^2$}\add{$W^2 = \sum_d (F_{\text{exact}}(d) - F_{\text{approx}}(d))^2$}~\cite{cramer1928composition}\del{ against exact CDFs}, were compared\del{ for $N, n \in \{2, \ldots, 500\}$}. \del{The beta distribution}\add{Beta}~\cite{johnson1995continuous} \del{achieves}\add{attains} the lowest \add{median }$W^2$\del{ across all pairs} (\del{Section~\ref{subsec:beta_quality}; Fig.~\ref{fig:dtv_N_50_n_10_and_beta}}\add{Fig.~\ref{fig:dtv}}), but \del{the second-ranked gamma distribution provides more conservative tail-area $p$-values that avoid overestimating rejection rates---e.g., at}\add{$W^2$ weights the whole support whereas a test depends only on the far right tail, where the second-ranked gamma is more accurate and conservative rather than anti-conservative: at} $N{=}500$, $n{=}256$\del{, beta reports power $0.399$ while gamma matches the exact $0.369$.}\add{ beta's critical DTV is $426$, not the exact $427$, inflating the ADC power of Section~\ref{subsubsec:adc} from $0.369$ to $0.383$; gamma reproduces both.} All \del{power comparisons}\add{approximate $p$-values} in Section~\ref{sec:results} therefore use \del{the gamma approximation with }$p \approx 1 - F_\gamma(d - 0.5)$, \del{applying the}\add{the} standard continuity correction for integer\del{-valued} statistics~\cite{wolfe2017intuitive}.

\subsection{Parameter estimation}
\label{subsec:estimation}

Distribution parameters are estimated by generating $K$ histograms under $H_0$ via \add{Monte Carlo (}MC\add{)} sampling~\cite{luengo2020survey}, computing DTV values, and fitting by minimizing the Cram\'{e}r--von Mises statistic. Convergence\add{ (beta family)} depends on $K/N$ but not on $n$ (\del{verified for $n \in \{5, 10, 20, 50\}$; Fig.~\ref{fig:mc_convergence}}\add{Fig.~\ref{fig:mc_convergence}, verified for $n \in \{5, 10, 20, 30\}$}). At $K/N \geq 100$ the mean absolute $p$-value error at $\alpha=0.05$ is within $2$\% (relative) of the exact-fit baseline\del{; at $K/N \geq 50$ within $3$\%. We use}\add{, so} $K = 50{,}000$\add{ is used}\del{, giving $K/N \geq 100$ for}\add{ for} $N \leq 500$; \del{for larger $N$, $K$ should be increased proportionally to maintain accuracy.}\add{larger $N$ needs $K$ raised proportionally.}

\section{Experimental results}
\label{sec:results}

\subsection{Beta approximation quality}
\label{subsec:beta_quality}

Fig.~\ref{fig:goodness_of_fit_beta} shows the Kolmogorov--Smirnov statistic between \del{exact and beta-approximated DTV distributions. Configurations with $n > N$ are included for theoretical completeness (Fig.~\ref{fig:heatmap_full}); in practice, histogram testing requires $N \geq n$. Restricting to $N, n \geq 100$ (Fig.~\ref{fig:heatmap_zoomed}) reveals that error decreases rapidly with $N$. For $N, n \in \{2, \ldots, 500\}$ ($249{,}001$ valid pairs), the critical DTV value at $\alpha=0.05$ was computed from both exact and beta-approximated CDFs at integer points. Critical values matched}\add{exact and beta-approximated DTV CDFs over the whole grid: it is large only in the degenerate $n>N$ region, decreases rapidly with $N$, and decays along the diagonal $N{=}n$ (Fig.~\ref{fig:heatmap_full}--\ref{fig:ks_diagonal}, respectively). Over all $249{,}001$ pairs $N, n \in \{2, \ldots, 500\}$, critical DTV values at the $\alpha=0.05$ level from exact and beta CDFs matched} exactly \del{in}\add{for} $50.7$\% \del{of cases }and differed by at most~$1$ \del{in}\add{for} $>99.9$\%\del{. The}\add{; the} median $p$-value error was $0.000116$ ($78.5$\% below $0.001$), \del{improving to}\add{or} $0.000089$ (\del{$97.1$\% below $0.001$) for $N \geq 5n$. Along the diagonal $N{=}n$, the KS error decays monotonically with increasing $N$ (Fig.~\ref{fig:ks_diagonal}).}\add{$97.1$\%) for $N \geq 5n$.}

\subsection{\add{Where CT is and is not sensitive}}
\label{subsec:where}

\add{No test is powerful against all alternatives; what matters is which it favors. Table~\ref{tab:l1} fixes $\Norm{\mathbf{p}-\mathbf{u}}{1}=\sum_i\A{p_i-1/n}=0.2$ ($\mathbf{u}$ uniform) and varies only the departure's \emph{shape}, at a moderate ($n{=}10$, $N{=}200$) and a sparse ($n{=}256$, $N{=}500$) operating point. Three order-aware competitors join chi-square and the $G$-test~\cite{chiang2003statistical}: the alternating contrast (AC) $\A{\sum_i(-1)^ix_i}$, i.e.\ Neyman's smooth test~\cite{neyman1937smooth} at the top frequency, hence locally most powerful against that comb; the discrete von Neumann ratio~\cite{von1941distribution} $\sum_i(x_i-x_{i-1})^2/\sum_i(x_i-\bar{x})^2$, the $\ell_2$ analogue of DTV; and the Wald--Wolfowitz runs test~\cite{wald1940test}, the number of sign runs of $x_i-N/n$: low for trends, high for alternation. Paninski's coincidence count (CO) $N-\A{\{i:x_i>0\}}$~\cite{paninski2008coincidence}, minimax-optimal in the sparse regime, is a modern baseline.}

\add{Three conclusions follow. First, DTV gains most in the period-2 (Nyquist) direction: there CT beats chi-square at both operating points and the $G$-test wherever the latter has a valid FPR, its edge decaying as the period grows. Second, it is least sensitive where the recent minimax literature's classes live: a monotone ramp, a single jump~\cite{lepski1999minimax}, and a depleted block~\cite{kipnis2026minimax} leave CT in the range $0.035$--$0.130$ while chi-square reaches $0.227$--$0.731$. Not dividing by expected cell counts, DTV handles information in few low-count cells less gracefully, so every advantage shrinks in the sparse column; there CO, blind to order but sensitive to occupancy, overtakes CT on half the shapes, led by the depleted block ($0.214$ vs.\ $0.036$), while trailing far behind on the comb ($0.123$ vs.\ $0.376$), so the two are complementary---and CO is vacuous once $N\gg n$. Third, the oracle AC, matched to the exact period and phase, beats CT there ($0.786$, $0.994$) but is erratic once the period changes, its fixed template aligning by luck or not ($0.040$ and $0.190$ at periods~4 and~6 for $n{=}10$, $0.043$ and $0.301$ for $n{=}256$), whereas CT is tuning-free and also fires on spiky and unstructured shapes. CT thus suits a suspected short-period disturbance of unknown period and phase---the ADC case below---while chi-square stays the default.}

\del{All power experiments use $50{,}000$ trials at $\alpha=0.05$ (SE~${<}\,0.003$ for all estimates). For comb distributions, CT has higher power than both chi-square and the $G$-test~\cite{chiang2003statistical}. To verify fairness, false positive rates (FPR) were measured ($n=10$, $N \in \{50, 100, 200, 500\}$): CT $0.043$--$0.048$, chi-square $0.044$--$0.049$, $G$-test $0.051$--$0.067$, MC-based CT $0.047$--$0.051$. The $G$-test's elevated FPR at small $N$ inflates its apparent power; exact CT is slightly conservative, so its advantage is unconfounded. CT is not a universal replacement: for non-comb deviations ($N{=}200$, $n{=}10$), chi-square retains far higher power (e.g., $0.918$ vs.\ $0.059$ for a monotonic trend); the tests are complementary. }CT is consistent \del{for}\add{against} comb alternatives: \add{under $H_0$ each $x_i\sim\text{Bin}(N,1/n)$, so adjacent differences are zero-mean with standard deviation $O(\sqrt{N/n})$, giving $\E{\text{DTV}}=O(\sqrt{nN})$ and, adjacent terms being weakly dependent, $\Var{\text{DTV}}=O(N)$; }under $H_1$ with $p_i = 1/n + (-1)^i\delta$\del{, $\E{\text{DTV}} = O(N)$; under $H_0$, $\E{\text{DTV}} = O(\sqrt{nN})$. Since $\Var{\text{DTV}} = O(N)$, the standardized statistic diverges, giving asymptotic power one.}\add{ each difference has mean $2\delta N$, so $\E{\text{DTV}}=2(n-1)\delta N=O(N)$. The standardized statistic grows like $O(\sqrt{N})$, so power $\to1$.}

\subsection{\del{Statistical power}\add{Applications}}
\label{sec:power}

\add{Unless noted: $50{,}000$ trials (SE~${<}\,0.003$), $\alpha{=}0.05$, $N{\geq}5n$.}

\subsubsection{Banker's rounding} \label{subsubsec:setup}

Round-half-to-even\del{ (banker's rounding)}~\cite{maxfield2006introduction}\add{, the IEEE~Std~754~\cite{ieee754} default and so pervasive in recorded data,} slightly favors even last digits, with per-digit probabilities $\frac{10^{r}+1}{10^{r+1}}$ (even) and $\frac{10^{r}-1}{10^{r+1}}$ (odd), where $r$ denotes the number of pre-rounding decimal places\del{, creating a comb-like distribution.}\add{, all removed: ties, of probability $10^{-r}$, go to the even neighbor, giving a period-2 comb of amplitude $\delta=10^{-(r+1)}$, CT's best direction.} Fig.~\ref{fig:mean_p_values} shows mean $p$-values over $10^6$ \del{samples}\add{histograms} per $N$; CT outperforms both \del{alternatives.}\add{competitors only for $N\gtrsim200$ ($r=1$) and $N\gtrsim2000$ ($r=2$), where $\delta^2$ is $100\times$ smaller.} Mean $p$-values are reported \del{because at moderate}\add{since at these} $N$ all tests have power barely above \del{$\alpha=0.05$, so rejection rates alone are uninformative at those sample sizes}\add{$\alpha$}.

\subsubsection{ADC differential nonlinearity}
\label{subsubsec:adc}

An analog-to-digital converter (ADC) maps \del{input}\add{its input} to one of $n$ codes\del{. Differential nonlinearity (DNL) quantifies bin width deviation: $\text{DNL}(k) = W(k)/W_{\text{ideal}} - 1$; IEEE Std~1241~\cite{ieee1241} checks uniformity via chi-square. SAR ADCs can exhibit \emph{alternating} DNL from capacitor mismatch~\cite{razavi2017principles} (the purely alternating model corresponds to LSB mismatch; higher-bit mismatch produces longer-period patterns). CT can supplement}\add{, and its differential nonlinearity $\text{DNL}(k)=W(k)/W_{\text{ideal}}-1$ measures how far the $k$-th code width $W(k)$ departs from the ideal; IEEE~Std~1241~\cite{ieee1241} tests it with a stimulus that would occupy all codes uniformly, running chi-square on the code histogram. In a successive-approximation ADC the code widths come from a binary-weighted capacitor array, so mismatch in the least significant capacitor widens every second code and narrows the rest~\cite{razavi2017principles}: $W_k=W_{\text{ideal}}(1+(-1)^k\varepsilon)$ with sampling probabilities $\propto 1\pm\varepsilon$, exactly the period-2 comb. Higher-order mismatch gives longer periods, where Table~\ref{tab:l1} shows CT loses its edge, so CT supplements rather than replaces} the IEEE~1241 chi-square step; applying both at $\alpha/2$\del{ each} controls the family-wise error\del{ rate}.

\del{An ADC with $n=10$ codes and alternating DNL of amplitude $\varepsilon$ was simulated (bin widths $W_k = W_{\text{ideal}}(1+(-1)^k\varepsilon)$, so sampling probabilities are $\propto 1 \pm \varepsilon$). At $N=200$, $\varepsilon=0.20$, CT achieves power $0.625$ vs.\ $0.452$ for chi-square ($38\%$ improvement) and $0.464$ for the $G$-test ($35\%$; Fig.~\ref{fig:adc_power}). MC-fitted CT closely tracks exact CT. At realistic resolution ($n=256$, 8-bit ADC, MC-gamma CT), $\varepsilon=0.20$, $N=500$: CT achieves power $0.369$ vs.\ $0.221$ for chi-square ($67\%$ improvement); the $G$-test is inapplicable as low expected counts ($N/n \approx 2$) produce catastrophic false positive rates ($0.19$--$0.54$ across $N \in \{200,\ldots,1000\}$). $N{=}500$ is deliberately small to accentuate power differences; practical ADC testing uses larger $N$. For i.i.d.\ random DNL ($n{=}10$, $\sigma=0.10$), all tests yield similar power (${\approx}\,0.10$); CT's gain is specific to alternating deviations. We recommend $N \geq 5n$ (gamma error ${<}\,0.001$ for $97.1\%$ of pairs per Section~\ref{subsec:beta_quality}).}\add{This re-runs the period-2 row of Table~\ref{tab:l1} with $\Norm{\mathbf{p}-\mathbf{u}}{1}=\varepsilon$, so ${\pm}0.009$ gaps are MC noise. For $n=10$, $N=200$, $\varepsilon=0.20$, CT reaches $0.625$ vs.\ $0.452$ ($\chi^2$) and $0.464$ ($G$) (Fig.~\ref{fig:adc_power}), the pre-fitted gamma reproducing the exact rejection rate; FPRs over $N\in\{50,\ldots,500\}$: CT $0.043$--$0.048$, MC-beta CT $0.047$--$0.051$, $G$ $0.051$--$0.067$. At a realistic resolution ($n=256$, 8-bit), $N=500$, $\varepsilon=0.20$, CT reaches $0.369$ vs.\ $0.221$, while $G$ is inapplicable: low expected counts ($N/n\approx2$) inflate its FPR to $0.19$--$0.54$ over $N\in\{200,\ldots,1000\}$.}

\subsubsection{\add{Real image data}} \label{subsubsec:images}

\add{A natural image's intensity histogram has no plausible uniform null; the \emph{output code} histogram of a quantizer driven by a real signal is well posed, and is exactly the IEEE~1241 setting. Nine 8-bit scikit-image test images~\cite{vandewalle2014scikit} were mapped to $[0,1)$ by the dithered rank transform a companding front end performs, which makes the ideal $n{=}256$ code distribution uniform up to one pixel per code, and requantized with the alternating widths $W_k$ above; drawing $N$ pixels at random gives a multinomial code histogram whose only departure from uniformity is $\varepsilon$---so these runs probe sample size and resolution, not image content. MC-gamma CT (moment-matched, so $K{=}50{,}000$ suffices at every $N$) keeps FPR at $0.046$--$0.048$ for $N\in\{500,2000,10^4,5\cdot10^4\}$ ($9{,}000$ trials each), two orders of magnitude beyond the largest exact $N$. At $N{=}10^4$, $\varepsilon{=}0.05$, CT reaches $0.493$ vs.\ $0.285$ ($\chi^2$) and $0.300$ ($G$); at $N{=}2000$, $\varepsilon{=}0.10$, $0.378$ vs.\ $0.225$ and $0.317$.}

\section{Conclusions}
\label{sec:conclusions}

\del{The comb test (CT) detects alternating histogram deviations with up to $67\%$ higher power than chi-square (at $n=256$, $N=500$, $\varepsilon=0.20$) at comparable FPR. A gamma approximation extends CT to large $N$ with conservative $p$-values. For non-comb alternatives chi-square is superior; the tests are complementary. Future work includes a two-sided variant, a circular DTV variant for cyclic data (e.g., hue histograms), and steganography detection, since LSB replacement embedding~\cite{fridrich2009steganography} induces comb patterns in pixel-value histograms.}\add{The comb test (CT) turns the discrete total variation of a histogram into a uniformity test with an exactly known null distribution, computed by dynamic programming for $N$ up to several hundred and, beyond that, by a gamma approximation whose FPR is verified here up to $N=5\cdot10^4$. Its useful domain is narrow but well defined: among $\ell_1$-matched alternatives it gains most against period-2 departures without needing their phase, degrades as the period grows, and is weak against smooth, single-jump, and depleted-block ones, where chi-square is superior. CT is therefore a targeted supplement to IEEE~1241. Future work includes a two-sided variant, a circular DTV, a non-uniform null via multinomial weights in~(\ref{eq:dp}), a bin-wise weighting restoring sensitivity in the sparse regime, and a minimax analysis of the comb direction~\cite{ingster2003nonparametric,kipnis2026minimax}. Scripts reproducing every number and figure are in the repository.}

\section*{Acknowledgment}

The authors thank Dr.\ Lenka Mihokovi\'{c} and Dr.\ Tomislav Buri\'{c} for discussions, and Herman Zvonimir Do{\v{s}}ilovi{\'{c}} for computational infrastructure. Claude Opus 4.6\add{ and~5} (Anthropic) \del{was used to assist with English}\add{assisted with} language editing\add{ and some experimental code}.

\bibliographystyle{IEEEtran}
\bibliography{literature}

\end{document}